\documentclass[a4paper, 10pt, aps, showkeys, showpacs, nofootinbib, twocolumn]{revtex4-2}

\usepackage{graphicx}
\usepackage{amsfonts}
\usepackage{amssymb}
\usepackage{amsbsy}
\usepackage{amsmath}
\usepackage{mathrsfs}
\usepackage{latexsym}
\usepackage{bm}
\usepackage{subfigure} 
\usepackage{wasysym}
\usepackage{mathbbol}
\usepackage{bigints} 
\usepackage{fontawesome}
\usepackage{natbib}
\usepackage{xcolor}
\usepackage{footnote}
\usepackage[colorlinks = true, linkcolor=blue, citecolor=cyan, urlcolor=blue]{hyperref}
\usepackage{orcidlink}
\usepackage{upgreek}
\usepackage{tabularx}

\begin{document}

\title[]{Effect of dark matter halo on transonic accretion flow around a galactic black hole}

\author{Subhankar Patra \orcidlink{0000-0001-7603-3923}}
\email{psubhankar@iitg.ac.in}

\author{Bibhas Ranjan Majhi \orcidlink{0000-0001-8621-1324}}
\email{bibhas.majhi@iitg.ac.in}

\affiliation{Department of Physics, Indian Institute of Technology Guwahati, Guwahati 781039, Assam, India}

\date{\today}

\begin{abstract}
We investigate the transonic accretion flow in the spacetime of a supermassive black hole (BH) coupled to an anisotropic dark matter fluid, as proposed by Cardoso {\it et al}. We essentially compare the accretion properties of the Cardoso BH with those of an isolated Schwarzschild BH. The Cardoso BH is described by the halo mass ($M_{\rm H}$) and its characteristic length scale ($a_0$). Various classes of accretion solution topologies (e.g., A and W-types) are obtained by solving the dynamical equations of the flow in a fully general relativistic framework. We find that the global accretion solutions in the identified solution topologies are substantially influenced by the halo parameters ($M_{\rm H}, a_0$) when the halo mass is high or the dark matter distribution is concentrated near the black hole. In this high compactness regime, different observational signatures of the accretion disc, like the spectral energy distribution (SED) and bolometric disc luminosity, are found to exhibit considerable deviations from the known results in the Schwarzschild BH model. Furthermore, we obtain shock-induced accretion solutions, where different shock properties, such as the shock radius ($r_{\rm sh}$), flow mass density ($\rho$) compression, and electron temperature ($T_e$) compression across the shock front, are potentially altered from those in the Schwarzschild BH model when the halo compactness is high. Interestingly, the existing shock parameter space, defined by the flow specific angular momentum ($\lambda$) and energy ($E$), is largely reduced for higher halo compactness compared to that of the Schwarzschild BH. These unique features offer a possible valuable tool for characterizing the presence or absence of a dark matter halo around a galactic black hole.
\end{abstract}




\maketitle
\section{Introduction}
\label{sec:introduction}

Astrophysical sources such as active galactic nuclei (AGN) and black hole X-ray binaries (BH-XRBs) are powered by the accretion of matter and emit electromagnetic radiation across all frequency domains \cite{Pringle-1981-137, Frank-2002-book}. Over timescales of a few days to months, their spectral state changes from the low-hard state (LHS) to the high-soft state (HSS) through several intermediate states \cite{Esin-1998-854, Nandi-2024-1149}. To understand their spectral characteristics, numerous accretion models have been proposed in the literature, depending on different physical conditions \cite[and references therein]{Abramowicz-2013-1, Yuan-2014-529}. Indeed, all those analyses provide essential properties of the accretion disc, such as mass accretion and outflow rates, size of the post-shock corona (PSC), disc inclination angle, quasi-periodic oscillation (QPO) frequency, photon index, etc \cite{Nandi-2018-1, Sreehari-2020-5891, Das-2021-2777, Sriram-2021-127, Majumder-2022-2508, Mondal-2022-A77, Heiland-2023-3834, Rawat-2023-113, Dhaka-2023-2721, Mondal-2024-A279, Nandi-2024-1149}. Also, these studies offer information about the central black holes, i.e., their mass and spin \cite{Molla-2016-88, Nandi-2018-1, Das-2021-2777, Mondal-2022-A77, Heiland-2023-3834, Mondal-2024-257}.

A number of scenarios support the presence of a dark matter halo around the supermassive black holes in AGNs \cite{Sadeghian-2013-063522}. Since dark matter interacts weakly, possibly through the weak nuclear force, its exact properties still remain unknown \cite{Bertone-2018-51}. However, it can interact gravitationally with the normal matter, thereby altering the geometry of spacetime. To understand how dark matter influences the gravitational wave (GW) and electromagnetic (EM) observations of black holes, several attempts have been made at the Newtonian level \cite{Eda-2013-221101, Macedo-2013-48, Barausse-2014-104059, Kavanagh-2020-083006}. However, to move beyond these approximate estimates, a proper spacetime manifold is required. Meanwhile, \citet{Cardoso-2022-L061501} proposed an exact analytical solution within Einstein's general relativity (GR), which represents a supermassive BH spacetime minimally coupled to a dark matter fluid with anisotropic pressure and Hernquist density profile. In this fully GR approach, few other BH metrics are modeled by considering different density profiles of the dark matter (e.g., King, Einasto, Jaffe, Burkert, Navarro–Frenk–White, Moore, Taylor–Silk profiles, etc.) \cite{King-1962-471, Einasto-1965-87, Jaffe-1983-995,  Burkert-1995-L25, Navarro-1996-563, Moore-1999-1147, DiCintio-2014-2986, Nishikawa-2017-043533, Xu-2018-038, Bhandari-2021-043, Konoplya-2022-166, Jusufi-2022-103, DeLuca-2023-048, Acharyya-2023-070, Liu-2024-731, Becar-2024-061, Gohain-2024-101683}. Although various properties of spacetime and phenomenology have been investigated to observe the influence of dark halos \cite{Retana-Montenegro-2012-A70, Jusufi-2020-354, Nampalliwar-2021-116, Igata-2022-2350105, Dai-2023-084080, Xavier-2023-064040, Myung-2024-03606, Kazempour-2024-043034, Heydari-Fard-2025-49, Chen-2024-172, Tan-2024-17760, Zhao-2024-06882, Mollicone-2024-024017, Pezzella-2024-18651, Amancio-2024-124048}, their effect on transonic matter accretion \cite{Liang-1980-271, Abramowicz-1981-314, Fukue-1987-309} onto black holes has not been studied. In this work, we aim to explore how the presence of a dark matter halo can be perceived through different transonic accretion properties. For this, we would like to concentrate on the Cardoso BH model only because, for most of the galaxies, it is consistent with the observed rotation curves and other dynamical properties as well. Needless to mention that the choice of the dark matter profile depends on the specific context and the kind of galaxy or dark matter halo being modeled.

The Cardoso BH is described by two independent parameters: halo mass ($M_{\rm H}$) and characteristic length scale ($a_0$). Recently, interest has grown in testing the Cardoso model on the rich astrophysical environments of the black holes, such as accretion discs, photon rings, etc. For example, in \cite{Cardoso-2022-L061501}, the authors investigated the effect of dark halo on GWs emission and propagation. The influence of dark matter components on the properties of EM radiations, like the quasi-normal modes, perturbations, scatterings, etc., has been studied in \cite{Konoplya-2021-136734}. The epicyclic oscillatory motion of the test particles and its application to the observed QPOs in AGN spectra were explored in \cite{Stuchlik-2021-059}. The investigation of the tidal forces and geodesic deviation motion due to a dark matter halo has been reported in \cite{Liu-2022-105104}. The evolution of the extreme mass ratio inspiral in a galactic black hole spacetime within the dark halo is analyzed in \cite{Dai-2023-084080}. In presence of a dark halo, the black hole shadow is studied in \cite{Xavier-2023-064040}, where the authors constrained the halo parameters ($M_{\rm H}$, $a_0$) using the Event Horizon Telescope (EHT) collaborations shadow data for the supermassive black holes $\text{M}87^{\star}$ and $\text{SgrA}^{\star}$. The impact of galactic environment on the geodesic motion of  extreme-mass-ratio binaries around a supermassive black hole, as well as their GW emission, has been studied in \cite{Destounis-2023-084027}. The effect of a dark matter halo on the motion of spinning particles was investigated in \cite{Tan-2024-17760}. The analysis of quasi-normal modes of a galactic black hole in a dark matter halo has been explored in \cite{Pezzella-2024-18651}. The energy spectrum and fluxes of the orbiting particles are examined in \cite{Heydari-Fard-2025-49} based on the Novikov-Thorne accretion model.

For black hole accretion, a key feature is that the flow must satisfy the inner boundary conditions at the event horizon. These conditions imply that the angular momentum of the flow should be sub-Keplerian near the horizon and cross the horizon at the speed of light. In that way, the flow motion must be transonic in nature, where the flow speed changes from subsonic to supersonic values \cite[and references therein]{Chakrabarti-1996-325, Kumar-2017-4221, Dihingia-2018-083004, Dihingia-2020-023012, Patra-2022-101120, Mitra-2022-5092, Sen-2022-048, Patra-2024-060, Mitra-2024-28, Patra-2024-371, Patra-2024-17108}. In the last few decades, the study of transonic accretion models has largely increased, as these models explain many observational signatures of the accretion disc near the black hole, such as hard power-law spectra, QPOs, and bipolar jets in the PSC, etc \cite{Das-2021-2777, Aktar-2018-17, Das-2022-1940, Dihingia-2019-2412, Patra-2024-371}. 

As we have seen, many strong gravity signatures have been analyzed in the Cardoso BH model, and some great results have been revealed. But, till now, to the best of our knowledge, nobody has reported the transonic accretion flows around the Cardoso BH. Such deficiency in the literature motivates us to serve the present work. We explore the global transonic accretion solutions and associated observational signatures (e.g., luminosity distributions, bolometric disc luminosity, etc.) in background of the Cardoso BH metric. Our results indicate that the halo parameters ($M_{\rm H}, a_0$) potentially affect the accretion disc properties when the dark matter compactness is high. However, for low compactness, these properties deviate insignificantly from the Schwarzchild BH. We compare the outcomes of the Cardoso BH model with those for the Schwarzchind BH model. We show that the high compactness of the halo largely modulates a given accretion solution topology (e.g., A and W-types) with respect to the usual Schwarzschild BH. Such effects change the flow temperature in the disc significantly. Consequently, the spectral energy distribution (SED) and bolometric disc luminosity ($L$) vary noticeably compared to the results in the Schwarzschild BH model. Moreover, we examine the shock solutions using relativistic shock conditions. We observe that for the Cardoso BH, shocks form at larger radii compared to the Schwarzschild BH. At high halo compactness, the shock radius ($r_{\rm sh}$) shifts significantly outward compared to that in the Schwarzschild BH model, leading to a potential decrease in mass density ($\rho$) compression and electron temperature ($T_e$) compression across $r_{\rm sh}$. In addition, we notice that the high compactness confines the shock solutions to a narrower range of flow specific angular momentum ($\lambda$) and energy ($E$) than in case of Schwarzschild BH. These feature may offer a potential tool for probing the presence of a dark matter environment around a galactic black hole.

The outlines of this paper are as follows. In Section \ref{sec:DM-metric}, we introduce the black hole metric with a dark matter halo. Section \ref{sec:model-equations} presents the governing flow equations for the accretion disc in a static and asymmetric spacetime. In Section \ref{sec:accretion-solutions}, we discuss the methodology used to find transonic accretion solutions and see the effect of the dark halo on solution topologies and their physical properties as well. In Section \ref{sec:shock}, we analyze the shock-induced accretion solutions and explore various shock properties as a function of halo compactness. The available parameter space for shocks and their modifications with halo parameters have been depicted in Section \ref{sec:shock-parameter-space}. Finally, in Section \ref{sec:conclusion}, we conclude our results.                

\section{Geometry of Galactic black hole with dark matter halo} 
\label{sec:DM-metric} 
In this section, we introduce the background spacetime, which has been used in our analysis, and discuss its properties. In \cite{Cardoso-2022-L061501}, the authors provided an exact analytical solution of Einstein's equations for describing a supermassive BH immersed in a dark matter halo. To do that, they follow the Einstein construction, where the anisotropic matter has tangential pressure only. They generalized the Einstein cluster, a technique to construct a stationary system of many gravitating masses, by including a black hole at the center of a dark matter distribution. Accordingly, the general relativistic geometry of such configuration is found to be \cite{Cardoso-2022-L061501},  
\begin{equation}
ds^2 = -f(r)dt^2 + \frac{dr^2}{1 - 2m(r)/r}
+  r^2d\Omega^2,
\label{eq:metric}
\end{equation}
where
\begin{equation}
	\begin{split}
		d\Omega^2 = d\theta^2 + \sin^2\theta d\phi^2.
	\end{split}
\end{equation}
The mass function $m(r)$ is chosen as, 
\begin{equation}
m(r) = M_{\text{BH}} + \frac{M_{\rm H}r^2}{(r + a_0)^2}\left(1 - \frac{2M_{\text{BH}}}{r}\right)^2,
\label{eq:mass-function}
\end{equation}
where $M_{\rm BH}$ is the mass of the central black hole, $M_{\rm H}$ is the mass of dark matter halo, and $a_0$ is the typical length scale that governs the size of the dark matter halo. The specialty of choosing such a mass profile is that it corresponds to the black hole mass $M_{\text{BH}}$ at small distances. On the other hand, at large scales, it describes the Hernquist density profile as,
 \begin{equation}
 	\label{eq:Hernquist-density-profile}
 	\begin{split}
 		\rho_0(r) = \frac{M_{\rm H}a_0}{2\pi r (r + a_0)^3}.
 	\end{split}
 \end{equation}
 
Using the mass profile (\ref{eq:mass-function}) and imposing the asymptotic flatness condition (i.e., $f \rightarrow 1$ at $r \rightarrow \infty$), the radial function is obtained from the Einstein's equation as, 
\begin{equation}
	\label{eq:f(r)}
f(r) = \left(1 - \frac{2M_{\text{BH}}}{r}\right)e^{\upgamma (r)},
\end{equation}
where
\begin{align}
	\upgamma (r) & = -\pi \sqrt{\frac{M_{\rm H}}{\xi}} + 2\sqrt{\frac{M_{\rm H}}{\xi}}\arctan\left(\frac{r + a_0 - M_{\rm H}}{\sqrt{M_{\rm H}\xi}}\right),\\
	\xi  & = 2a_0 - M_{\rm H} + 4M_{\rm{BH}}.
\end{align}

The matter density ($\rho_0$) and tangential pressure ($P_t$) corresponding to the solution (\ref{eq:f(r)}) are obtained as,
\begin{align}
	\label{eq:density-profile}
		\rho_0(r) &  = \frac{1}{4\pi r^2}\frac{dm(r)}{dr} = \frac{M_{\rm H}\left(a_0 + 2M_{\text{BH}}\right)\left(1 - 2M_{\text{BH}}/r\right)}{2\pi r(r + a_0)^3},\\
		P_t (r) & = \frac{m(r)\rho_0}{2[r - 2m(r)]}.
\end{align}

The black hole solution described by Eq.~(\ref{eq:metric}) has a regular event horizon at $r = r_{\rm H} = 2M_{\text{BH}}$ and the curvature singularity at $r = 0$. In the asymptotically flat regime (i.e., $r \rightarrow \infty$), the mass of the spacetime (\ref{eq:metric}) is referred to as the Arnowitt-Deser-Misner (ADM) mass, which is given by \cite{Arnowitt-1961-997, Cardoso-2022-L061501},
	\begin{equation}
		\label{eq:ADM-mass}
		M_{\text{ADM}} = \lim\limits_{r  \rightarrow \infty} m(r) = M_{\text{BH}} + M_{\rm H}.
\end{equation}
It is noted that at the horizon, $\rho_0$ vanishes, while $P_t$ remains regular. Moreover, the dark matter fluid satisfies both the weak and strong energy conditions everywhere outside the horizon, as both $\rho_0$ and $P_t$ are always positive. However, near $r_{\rm H}$, $P_t/\rho_0$ diverges because $\rho_0$ becomes very small. As a result, the dominant energy condition is violated in this region. Nevertheless, this does not affect the spacetime dynamics as the near-horizon region is nearly empty due to the small values of $\rho_0$ and $P_t$. Here, the scale hierarchy $M_{\text{BH}} << M_{\rm H} << a_0$ provides the relevant astrophysical setup\footnote{Astrophysical setup is featured by the absence of any curvature singularities outside the black hole event horizon ($r_{\rm H}$). Therefore, in the region of spacetime accessible to the external observer, i.e., $r > r_{\rm H}$, the spacetime must remain regular.} (for more detail see \cite{Cardoso-2022-L061501, Konoplya-2021-136734, Stuchlik-2021-059, Xavier-2023-064040, Heydari-Fard-2025-49}). In this work, we focus only on cases where the spacetime parameters fulfill the above mentioned inequities. To quantify the compactness of the halo, we introduce a parameter called compactness parameter, defined as $C = M_{\rm H}/a_0$.
 
 \section{Model equations governing accretion disc}
 \label{sec:model-equations}
The dynamical equations governing the accretion flow in the spacetime (\ref{eq:metric}) have been developed in this section. We model the hydrodynamics of accretion flow within a complete general relativistic setting \cite{Rezzolla-2013-book}. We assume that the motion of an ideal fluid is confined to the equatorial plane (i.e., $\theta = \pi/2$) of the central black hole, meaning the flow has no transverse motion (i.e., $u^{\theta} = 0$, where $u^{\theta}$ is $\theta$ component of the four-velocity $u^{k}$). Also, we have $\partial_{\theta}Q = 0$, where $Q$ is any flow parameter (e.g., mass density, pressure, and temperature, etc.). Moreover, the fluid is steady (i.e., $\partial_t Q = 0$) and obeys the azimuthal symmetry of the spacetime (i.e., $\partial_{\phi}Q = 0$). To simplify the fluid motion to one dimension (radial motion only), we adopt a co-rotating frame (CRF), which rotates with the same angular velocity as the fluid. In this work, we choose a unit system such that $G = M_{\text{BH}} = c = 1$, where $G$ is the gravitational constant and $c$ is the speed of light. Such a choice makes all the physical quantities dimensionless. Under these assumptions, the radial momentum equation can be written as \cite{Dihingia-2018-083004},
\begin{equation}
	\label{eq:radial-momentum}
	\gamma_{v}^{2}v\frac{dv}{dr} + \frac{1}{e+p}\frac{dp}{dr} + \left(\frac{d\Phi^{\text{eff}}}{dr}\right)_\lambda = 0,
\end{equation}
where $\gamma_{v}$ is the Lorentz-factor corresponding to the radial component of the physical three-velocity ($v$) in the CRF, $e$ is the total internal energy density, $p$ is the isotropic fluid pressure, $\Phi^{\rm eff}$ is the effective potential of the system, and $\lambda$ ($= - u_{\phi}/u_{t}$, where $u_{\phi}$ and $u_{t}$ are the $\phi$ and $t$ components of $u_k$) is the specific angular momentum of fluid. Note that for accretion, $v$ is a negative quantity. The expression of $\Phi^{\rm eff}$ is obtained in terms flow parameter $\lambda$ and spacetime parameters ($M_{\rm H}, a_0$) as,
\begin{equation}
	\label{eq:eff-potential}
	\Phi^{\rm eff} = 1 + \frac{1}{2}\ln(\Phi),~\Phi = \frac{r^2(r-2)e^{\upgamma (r)}}{r^3 - \lambda^2(r-2)e^{\upgamma (r)}}.
\end{equation}

In steady state, mass accretion rate ($\dot{M}$) is usually taken as a constant of motion (i.e., $d\dot{M}/dr = 0$). Integrating the conservation equation of mass flux (i.e., $\nabla_k(\rho u^{k}) = 0$, $\rho$ is the mass density of flow), we get the expression of $\dot{M}$ as,
\begin{equation}
	\label{eq:mass-accretion-rate}
	\dot{M} = -4\pi \rho H v \gamma_{v} \sqrt{r(r-2)e^{\upgamma (r)}} = {\text{constant}},
\end{equation}
where $H$ is the half-thickness of the disc. Considering the hydrostatic equilibrium along the vertical direction of the disc, $H$ is calculated as \cite{Lasota-1994-341, Riffert-1995-508, Peitz-1997-681},
\begin{equation}
	\label{eq:half-thickness}
	H = \sqrt{\frac{pr^3}{\rho F}},
\end{equation}
where $F = 1/(1 - \Omega\lambda)$. The angular velocity ($\Omega$) of the flow is given by, 
\begin{equation}
	\label{eq:angular-velocity}
	\Omega = \frac{u^{\phi}}{u^{t}} = \frac{\lambda (r-2)e^{\upgamma (r)}}{r^3}.
\end{equation}

In our model, the other constant of motions can be found from the present spacetime symmetries. We find two conserved quantities along the streamlines of the flow as (a) Bernoulli constant: $E = - (e + p)u_t/\rho$ (from time-translation symmetry), and (b) bulk angular momentum: $\mathcal{L} = (e + p)u_\phi/\rho$ (from azimuthal symmetry). Therefore, $\lambda$ ($= - \mathcal{L}/E$) appears to be another constant of motion.

We consider a relativistic equation of state as proposed in \cite{Chattopadhyay-2009-492}, where a variable adiabatic index $\Gamma$ is used instead of assuming a constant value. Following that work, the thermodynamic variables $e$ and $p$ can be found as,
\begin{equation}
	\label{eq:EoS}
	e = \frac{\rho f}{1 + m_p/m_e}, \hspace{0.25cm} p = \frac{2\rho\Theta}{1 + m_p/m_e}, 
\end{equation}
where $m_p$ is the proton mass and $m_e$ is the electron mass. The quantity $f$ is expressed in term of dimensionless temperature $\Theta$ ($= k_{B}T/(m_{e}c^{2})$, $k_{B}$ is the Boltzmann constant and $T$ is the flow temperature in Kelvin) as,
\begin{equation}
	\label{eq:f}
	f = 1 + \frac{m_p}{m_e} + \Theta\left[\frac{9\Theta + 3}{3\Theta + 2} + \frac{9\Theta + 3m_{p}/m_{e}}{3\Theta + 2m_{p}/m_{e}}\right].
\end{equation}

After solving the equation $d\dot{M}/dr = 0$ using Eqs. (\ref{eq:mass-accretion-rate}), (\ref{eq:half-thickness}), and (\ref{eq:EoS}), the temperature gradient of the flow is obtained as,
\begin{equation}
	\label{eq:temperature-gradient}
	\frac{d\Theta}{dr} = - \frac{2\Theta}{\left(df/d\Theta\right) + 1}\left(\frac{\gamma_v^2}{v}\frac{dv}{dr} + N_{11} + N_{12}\right),
\end{equation}
with
\begin{equation}
	N_{11} = \frac{3}{2r} + \frac{r-1}{r(r-2)} - \frac{1}{2F}\frac{dF}{dr},~N_{12} = \frac{1}{2}\frac{d\upgamma}{dr}.
\end{equation}

As $\dot{M}$ is very small for the context of supermassive black hole accretion \cite{Yuan-2014-529, Yarza-2020-50}, we neglect the radiative cooling mechanism in the energy equation (first law of thermodynamics). Therefore, it is obtained as,
\begin{equation}
	\label{eq:energy-equation}
	\frac{e+p}{\rho}\frac{d\rho}{dr} - \frac{de}{dr} = 0.
\end{equation} 

The expression of $\rho$ is calculated by integrating Eq. (\ref{eq:energy-equation}) as,
\begin{equation}
	\label{eq:mass-density}
	\rho = \mathcal{K}e^{\chi}\Theta^{3/2}(3\Theta + 2)^{3/4}(3\Theta+2m_p/m_e)^{3/4}, 
\end{equation}
where $\mathcal{K}$ refers the entropy constant and $\chi = (f - 1 - m_p/m_e)/(2\Theta)$. From Eq.~(\ref{eq:energy-equation}), it is evident that the flow is locally adiabatic, which implies constant entropy content. Following the works of \cite{Chattopadhyay-2016-3792, Kumar-2017-4221}, the entropy accretion rate of the flow is found to be,
\begin{equation}
	\label{eq:entropy-accretion-rate}
	\begin{split}
		\mathcal{\dot{M}} = \frac{\dot{ M}}{4\pi\mathcal{K}} & = - v\gamma_{v}H\sqrt{r(r-2)e^{\upgamma (r)}}\\
		& \times e^{\chi}\Theta^{3/2}(3\Theta + 2)^{3/4}(3\Theta+2m_p/m_e)^{3/4}.
	\end{split}
\end{equation}

To obtain the radial velocity gradient, we simultaneously solve Eqs.~(\ref{eq:radial-momentum}), (\ref{eq:EoS}), (\ref{eq:temperature-gradient}) and (\ref{eq:energy-equation}), which leads to the result,
\begin{equation}
	\label{eq:velocity-gradient}
	\frac{dv}{dr} = \frac{\mathcal{N}}{\mathcal{D}},
\end{equation}
The expressions of numerator ($\mathcal{N}$) and denominator ($\mathcal{D}$) of the above equation are found to be,
\begin{align}
	\mathcal{N} & = \frac{2C_{s}^2}{\Gamma + 1} (N_{11} + N_{12}) - \frac{d\Phi^{\rm eff}}{dr},\\
	\mathcal{D} & = \gamma_{v}^{2}\left[v - \frac{2C_{s}^{2}}{(\Gamma + 1)v}\right],
\end{align}
where $C_s$ ($= \Gamma p/(e+p)$) is the adiabatic sound speed with $\Gamma = 1 + 2/(df/d\Theta)$.

We consider the emission of thermal bremsstrahlung radiation from the accretion disc. Since the disc medium is optically thin for the hot accretion flow (HAF) \cite{Yuan-2014-529, Patra-2024-060}, bremsstrahlung radiation can escape from the disc without being absorbed \cite{Patra-2024-060}. We assume a completely ionized hydrogen plasma (atomic number $Z = 1$), where the number densities of electrons and ions are the same, i.e., $n_e = n_p \approx \rho/m_p$. Moreover, we use an approximate expression for the free-free emission coefficient, as proposed by \citet{Novikov-1973-343}, given by,
\begin{equation}
	\label{eq:emissivity}
	\begin{split}
		\mathcal{E}_{\nu_e}^{\rm ff} & = 6.8\times 10^{-38}(\rho/m_p)^2T_{e}^{-1/2}(1 + 4.4\times 10^{-10}T_{e})
		\\& \times \exp\biggl(-\frac{h\nu_e}{k_{B}T_{e}}\biggr)\bar{g}_{\text{ff}}~{\text{erg}~\text{s}^{-1}~\text{cm}^{-3}~\text{Hz}^{-1}},
	\end{split}
\end{equation} 
where $h$ is the Planck constant, $T_e$ is the electron temperature, $\nu_e$ is the emission frequency, and $\bar{g}_{\rm ff}$ is the thermally-averaged Gaunt factor (which includes quantum mechanical correction). In our analysis, we take $\bar{g}_{\rm ff} = 1.2$ \cite{Yarza-2020-50}. The second term in Eq.~(\ref{eq:emissivity}) accounts for both electron-electron emission and relativistic corrections. It is important to note that for HAF, their effectiveness is significant \cite{Patra-2024-060}. In the accretion disc, since the in-fall timescale is shorter than the ion-electron collision timescale, it is challenging to maintain thermal equilibrium between the ions and electrons. Typically, the electron temperature is lower compared to the ion temperature because the electron mass is much smaller than the mass of the ions. Several studies in the literature have self-consistently calculated the electron and ion temperatures \cite{Dihingia-2018-1, Dihingia-2018-2164, Dihingia-2020-3043, Sarkar-2022-34}. However, for simplicity, various scaling relations between $T_e$ and $T$ \cite{Dihingia-2018-1, Dihingia-2020-023012, Sen-2022-048} are often used, with $T$ calculated self-consistently by solving Eq. (\ref{eq:temperature-gradient}). In this study, we also adopt a scaling relation $T_e = T/10$ to remain consistent with the work of \cite{Yarza-2020-50}.

For an observer at spatial infinity, the emission frequency is redshifted due to the strong gravitational potential of the central black hole, as well as the rotation of the disc. For simplicity, we neglect any light-bending effects on the emitted radiation. Additionally, the velocity distribution of the electrons is assumed to follow the standard Maxwell's prescription. Under these assumptions, the red-shift factor ($1 + z$) is found to be \cite{Luminet-1979-228, Rybicki-1991-book, Sen-2022-048},
 \begin{equation}
	\label{eq:red-shift}
	1 + z = \frac{\nu_e}{\nu_o} = u^{t}\left(1 + \frac{r\Omega}{c}\sin{\theta_0}\sin{\phi}\right),	
\end{equation}
with 
\begin{equation}
	u^t = \gamma_v \sqrt{\frac{r}{(1 - \Omega \lambda)(r-2)e^{\upgamma (r)}}}.
\end{equation}
Here, $\nu_o$ is the observed frequency and $\theta_0$ is the inclination angle of the accretion disc with respect to the distant observer frame. We take $\theta_0 = 45^{\circ}$ for the purpose of illustration. Using Eqs.~(\ref{eq:emissivity}) and (\ref{eq:red-shift}), we get the monochromatic disc luminosity measured by an observer at infinity as,
\begin{equation}
	\label{eq:monochromatic-luminosity}
	L_{\nu_o} = 2\int_{r_{\rm H}}^{r_{\rm edge}}\int_{0}^{2\pi}\mathcal{E}_{\nu_{o}}^{\rm ff}Hrdr d\phi~{\text{erg}~\text{s}^{-1}~\text{Hz}^{-1}},
\end{equation}
where $r_{\rm H}$ is taken as the inner edge of the disc. In this analysis, the outer edge of the disc is assumed to be at $r_{\text{edge}} = 1000$.

Finally, integrating Eq.~(\ref{eq:monochromatic-luminosity}) over all frequency domains, we calculate the bolometric disc luminosity as,
\begin{equation}
	\begin{split}
		\label{eq:bolometric-luminosity}
		L & = \int_{0}^{\infty}L_{\nu_o}d\nu_{o} 
		\\& = 6.8\times 10^{-38}\left(\frac{2k_B}{hm_p^2}\right)\bar{g}_{\rm ff}~{\rm erg~s^{-1}} 
		\\& \times \int_{r_{\rm H}}^{r_{\rm edge}}\int_{0}^{2\pi}\frac{\rho^{2}T_{e}^{1/2}(1 + 4.4\times 10^{-10} T_{e})Hr}{u^{t}\left(1 + \frac{r \Omega \sin{\phi}}{\sqrt{2}c}\right)}~drd\phi.
	\end{split} 
\end{equation}
The above equations are useful for finding the accretion solutions and their corresponding disc properties, such as temperature profile, disc luminosity, and spectral energy distribution, etc. Note that when we set halo mass $M_{\rm H} = 0$ in these equations, we can recover the flow equations in the usual Schwarzschild BH spacetime. A detailed discussion of the accretion properties around the galactic black hole metric (\ref{eq:metric}) is provided in Section \ref{sec:results}.

\section{Results}
\label{sec:results}

\begin{figure}
	\centering
	\includegraphics[width=\columnwidth]{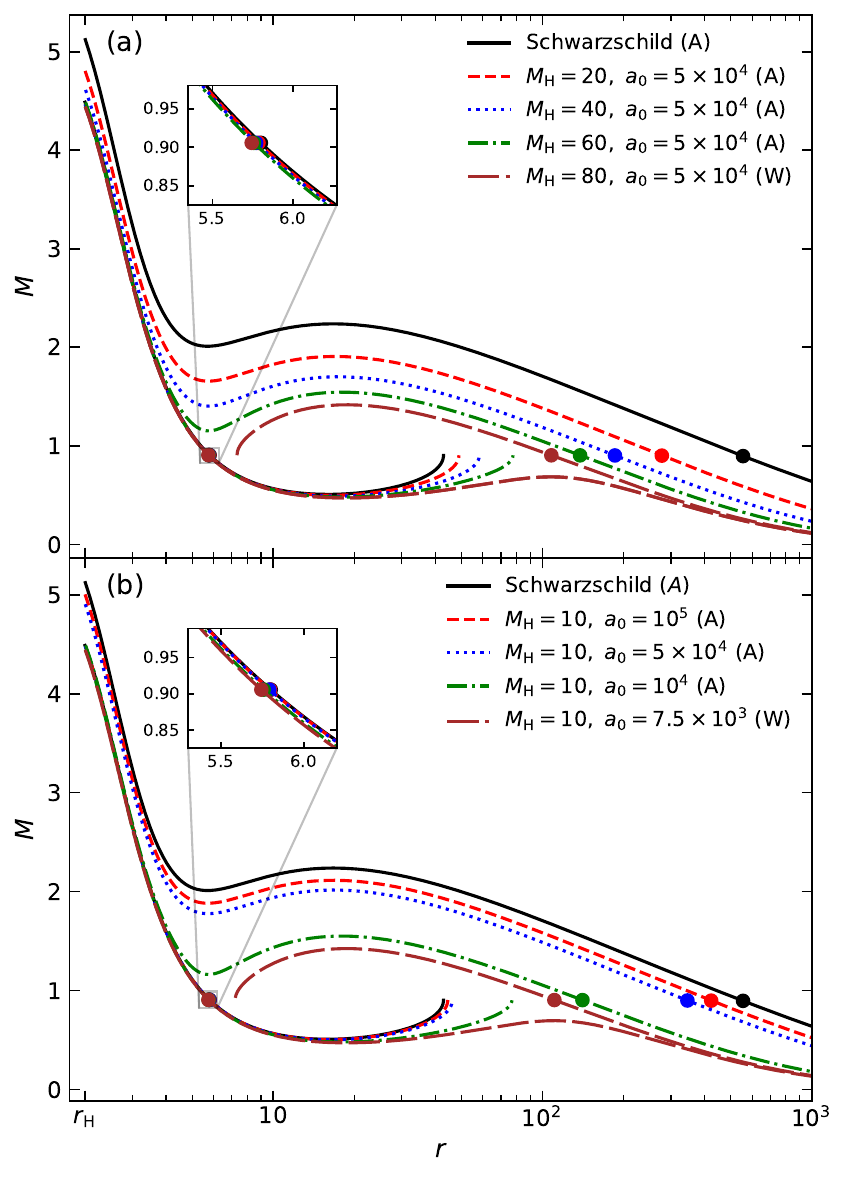}
	\caption{Typical accretion solutions (i.e., Mach number ($M = |v|/C_s$) versus radial distance ($r$) plots) for halo masses $M_{\rm H} = 20$, $40$, $60$, and $80$ with length scale $a_0 = 5 \times 10^4$ (panel (a)), and for $a_0 = 10^5$, $5 \times 10^4$, $10^4$, and $7.5 \times 10^3$ with $M_{\rm H} = 10$ (panel (b)). In each panel, the solid (black) line corresponds to the Schwarzschild black hole without a dark matter halo. The critical points are marked by filled circles. The types of solution topologies are indicated in each panel. In this figure, we choose $\lambda = 3.1$ and $E = 1.00025$. See the text for details.}
	\label{fig:solutions}
\end{figure}

\subsection{Transonic accretion solutions}
\label{sec:accretion-solutions}
This section explores the transonic accretion solutions, where the flow must pass through at least one critical point \cite{Liang-1980-271, Abramowicz-1981-314}. Critical points ($r_c$) are such radial coordinates where the velocity gradient $dv/dr$ (see Eq. (\ref{eq:velocity-gradient})) takes the form $``0/0"$. Therefore, the necessary conditions for finding the critical points are $\mathcal{N} = \mathcal{D} = 0$. Note that flow may possess single or multiple critical points depending on the global constants $\lambda$, $E$, $M_{\rm H}$, and $a_0$. The critical points that are formed close to the horizon are called inner critical points ($r_{\text{in}}$), and those formed far away from the horizon are called the outer critical points ($r_{\text{out}}$). As $(dv/dr)_{r_c}$ takes an indeterminate form, we use the l$'$H\^{o}pital's rule to Eq. (\ref{eq:velocity-gradient}) for finding the finite values of $dv/dr$. Usually, $(dv/dr)_{r_c}$ have two values. Depending on them, critical points are classified into three categories --- (a) saddle-type: $(dv/dr)_{r_c}$ values are real with opposite sign; (b) nodal-type: $(dv/dr)_{r_c}$ values are real with the same sign; and (c) spiral-type: both values of $(dv/dr)_{r_c}$ are imaginary. The positive value of $dv/dr$ corresponds to an accretion solution, and the negative value of $dv/dr$ yields a wind solution. Therefore, out of three types of critical points, only the saddle-type critical points (hereafter called critical points) are physically acceptable. In this work, we focus on accretion solutions that only pass through saddle-type critical points, excluding any analysis of wind solutions. To find the accretion solutions, we first calculate $r_c$ and its corresponding flow variables $\Theta_c$ and $v_c$ for a given set of global constants ($\lambda, E, M_{\rm H}, a_0$). Using those results as an initial boundary condition, we then numerically solve the differential Eqs.~(\ref{eq:temperature-gradient}) and (\ref{eq:velocity-gradient}) form $r_c$ to $r_{\text{edge}}$ and also from $r_c$ to $r_{\rm H}$. Finally, combining the two segments of the solution, we obtain a complete accretion solution.

Following the above methodology, we find the transonic accretion solutions for different sets of input parameters. The obtained results are presented in Fig.~\ref{fig:solutions}a, where the Mach number ($M = |v|/C_{s}$) is plotted as a function of the radial distance ($r$). Here, the flow parameters are chosen as $\lambda = 3.1$ and $E = 1.00025$. We fix the length scale at $a_0 = 5 \times 10^4$ and vary the halo mass $M_{\rm H}$. The solid (black) curve corresponds to the Schwarzschild BH without dark matter halo (i.e., $M_{\rm H} = 0$), while the dashed (red), dotted (blue), dash-dotted (green), and long-dashed (brown) curves represent the results for the Cardoso BH with $M_{\rm H} = 20$, $40$, $60$, and $80$, respectively. For the Schwarzschild BH model, flow is found be posses multiple critical points at $r_{\text{in}} = 5.8001$ and $r_{\text{out}} = 554.8449$. We notice that the solution passing through $r_{\text{out}}$ extends from $r_{\rm edge}$ to $r_{\rm H}$ (i.e., an open or global solution). However, the solution passing through $r_{\text{in}}$ is unable to connect $r_{\rm edge}$ and $r_{\rm H}$ (i.e., a closed solution) and terminates at the radius $r_t = 42.8606$. In the transonic accretion model, the flow begins its journey from the disc outer edge and then passes through a critical point. Thereafter, it continues to proceed toward the central black hole until it reaches the event horizon. This behavior is required in order to satisfy the inner boundary condition at the horizon (i.e., $v$ approaches $c$ at $r_{\rm H}$). Only such solutions, which extend continuously from the disc’s outer edge to the horizon, can be considered physically acceptable. Therefore, we find that a topology in which the solution through $r_{\rm out}$ is physically acceptable, whereas that through $r_{\rm in}$ is not. Such a topology is referred to as the A-type solution topology. Now, in the presence of a dark matter halo with $M_{\rm H} = 20$, the flow continues to exhibit A-type solution topology. However, in this case, the global solution passes through the outer critical point at a relatively smaller radius, $r_{\rm out} = 277.2015$, compared to the Schwarzschild BH case. And, the closed solution passing through $r_{\rm in} = 5.7853$ terminates at a larger radius, $r_t = 48.8953$. Interesting to know, A-type solution topologies can generate shock waves \cite{Chakrabarti-1989-365, Dihingia-2019-2412, Patra-2022-101120, Sen-2022-048, Patra-2024-060, Mitra-2024-28, Patra-2024-371}, which are extensively discussed in Section \ref{sec:shock}. When $M_{\rm H}$ is increased to $40$ and $60$, the solution topologies still remain A-type, where the critical points are formed at even smaller radii, $(r_{\text{in}}, r_{\text{out}}) = (5.7709, 185.5785)$ and $(5.7569, 137.7683)$, respectively. The termination radii for the inner critical point solutions are found to be even larger, at $r_t = 58.2309$ and $77.6369$, respectively. With a further increase in the halo mass to $M_{\rm H} = 80$, the flow still possesses multiple critical points $(r_{\rm in} = 5.7431,\, r_{\rm out} = 107.7519)$, but they are located relatively closer to the horizon. In this case, the solution passing through $r_{\rm out}$ is found be a closed one, while the solution through $r_{\rm in}$ is a global one. Such behavior of the accretion solutions in a topology is referred to as the W-type solution topology. We calculate the termination radius for the outer critical point solution as $r_t = 7.3319$. Similarly, in Fig. \ref{fig:solutions}b, we present the different accretion solution topology for a fixed halo mass $M_{\rm H} = 10$ with varying length scale $a_0$. For this, we choose the same set of flow parameters ($\lambda, E$) as used in Fig.~\ref{fig:solutions}a. The obtained results are shown using the dashed (red), dotted (blue), dash-dotted (green), and long-dashed (brown) lines for $a_0 = 10^5$, $5 \times 10^4$, $10^4$, and $7.5 \times 10^3$, respectively. Here, the solid (black) curve corresponds to the usual Schwarzschild BH case, as shown in Fig. \ref{fig:solutions}a, and has been included again for the comparison with the results for dark matter model. For Cardoso BH with large values of $a_0$, such as $10^5$, $5 \times 10^4$, and $10^4$, the solution topologies remain A-type, similar to the Schwarzschild BH, with the critical points at $(r_{\rm in}, r_{\rm out}) = (5.7956, 422.6707)$, ($5.7912, 344.8331$), and ($5.7569, 140.5995$), respectively. For a smaller value of $a_0$, such as $7.5 \times 10^3$, the solution topology changes to W-type from A-type, where the critical points are calculated as $(r_{\rm in}, r_{\rm out}) = (5.7432, 110.5782)$. For panel (b), the corresponding termination radii are calculated as $r_t = 44.4556$, $46.2112$, $76.9269$, and $7.2282$. It is noted that for both A and W-type topologies, only open (or global) solutions are usable for carrying out a physically relevant analysis, as they satisfy the necessary conditions for transonic accretion flows. In contrast, closed solutions are discarded, as they fail to represent a complete and physically consistent flow structure. However, closed solutions in A-type topologies become relevant to accretion dynamics only when shocks are taken into account \cite[and references therein]{Dihingia-2019-2412, Sen-2022-048, Mitra-2024-28}. The mechanism by which such closed solutions in A-type topologies give rise to shocks is thoroughly discussed in Section \ref{sec:shock}. These shocks can account for the observed QPOs in the luminosity spectrum of the accretion disc \cite{molteni-1995, Chakrabarti-2015}. Moreover, the post-shock flow can generate high-energy X-ray radiation, which is usually detected in the spectra of AGNs \cite{Majumder-2022-2508, Nandi-2024-1149, Chatterjee-2024-148}. Further discussions on QPOs and high-energy X-ray emission originating from the post-shock disc are presented in Section \ref{sec:shock} as well. Therefore, both A and W-type solution topologies are astrophysically significant, as they help to explain various observational features of transonic accretion flows around black holes.

\begin{figure*}
	\centering
	\includegraphics[width=0.8\linewidth]{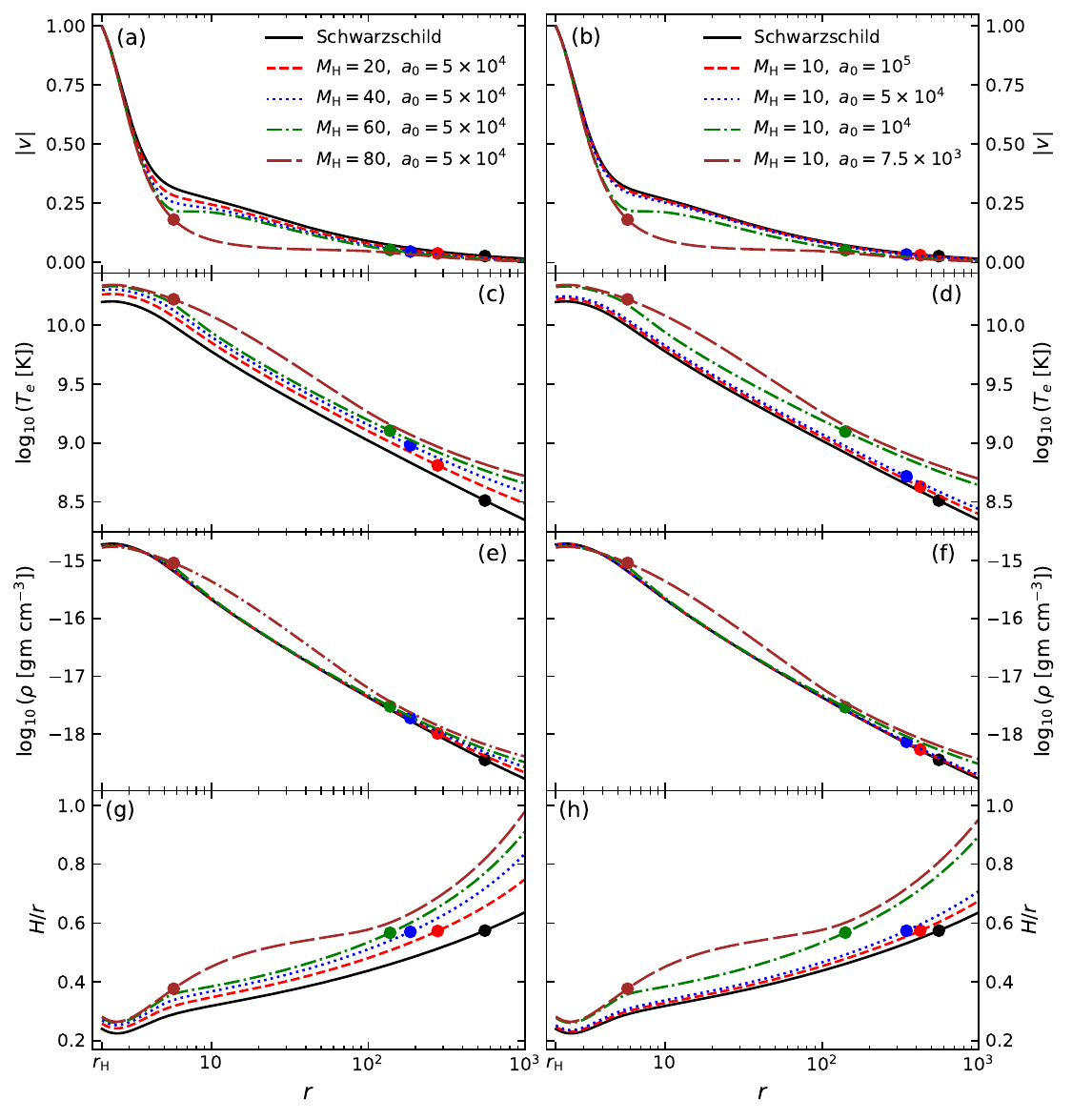}
	\caption{Profiles of flow variables corresponding to the global accretion solutions presented in Fig. \ref{fig:solutions}. The radial velocity ($v$), electron temperature ($T_e$), mass density ($\rho$), and aspect ratio ($H/r$) profiles for Figs. \ref{fig:solutions}a-b are shown in panels (a)-(b), (c)-(d), (e)-(f), and (g)-(h), respectively. See the text for details.}
	\label{fig:fp}
\end{figure*}

\begin{table*}
	\centering
	\caption{Dark matter halo mass ($M_{\rm H}$), halo length scale ($a_0$), critical point locations ($r_{\text{in}}, r_{\text{out}}$), critical point temperatures ($T(r_{\text{in}}), T(r_{\text{out}})$), and topology types for the accretion solutions presented in Fig.~\ref{fig:solutions}.}
	\label{tab:table-1}
	\resizebox{\linewidth}{!}{%
	\begin{ruledtabular}
		\begin{tabular}{lcccccc}
			$M_{\rm H}$ & $a_0$ & $r_{\text{in}}$ & $r_{\text{out}}$ & $T(r_{\text{in}})$ &$T(r_{\text{out}})$ & $\text{Type}$\\
			& & & & ($\times 10^{10}~{\text{K}}$) & ($\times 10^{9}~{\text{K}}$)& \\
			\hline
			$0$ (Schwarzschild) & --- & $5.8001$ & $554.8449$ & $1.6101$ & $0.3242$ & A\\
			$20$ & $5 \times 10^4$ & $5.7853$ & $277.2015$ & $1.6215$ & $0.6459$ & A\\
			$40$ & $5 \times 10^4$& $5.7709$ & $185.5785$ & $1.6328$ & $0.9554$ & A\\
			$60$ & $5 \times 10^4$& $5.7569$ & $137.7683$ & $1.6441$ & $1.2719$ & A\\
			$80$ &$5 \times 10^4$ & $5.7431$ & $107.7519$ & $1.6618$ & $1.5982$ & W\\
			$10$ & $10^5$ & $5.7956$ & $422.6707$ & $1.6135$ & $0.4254$ & A\\
			$10$ & $5 \times 10^4$ & $5.7912$ & $344.8331$ & $1.6169$ & $0.5208$ & A\\
			$10$& $10^4$ & $5.7569$ & $140.5995$ & $1.6440$ & $1.2499$ & A\\
			$10$& $7.5 \times 10^3$ & $5.7432$ & $110.5782$ & $1.6617$ & $1.5636$ & W\\
		\end{tabular}
	\end{ruledtabular}
}
\end{table*}
The radial velocity profiles corresponding to the global accretion solutions of panels (a) and (b) of Fig. \ref{fig:solutions} are shown in Figs.~\ref{fig:fp}a-b, respectively. We observe that the flow velocity is minimal (i.e., $v << 1$) at the outer region of the disc. As the flow moves towards the black hole, $v$ increases and eventually exceeds the local sound speed $C_s$ after passing through the critical point $r_c$. Subsequently, the flow becomes supersonic and continues to move towards the horizon. Finally, at $r_{\rm H}$, $v$ approaches the light speed $c$, satisfying the inner boundary condition of the transonic accretion model. In Figs.~\ref{fig:fp}c-d, we present the respective profiles of the electron temperature ($T_e$) for the global accretion solutions shown in Figs.~\ref{fig:solutions}a-b. In all cases, $T_e$ increases as we move towards $r_{\rm H}$ from $r_{\text{edge}}$. We observe that the temperature distribution of the disc rises as $r_c$ drifts toward the horizon. Also, the solutions associated with $r_{\text{in}}$ exhibit relatively higher $T_e$ profiles compared to the solutions that pass through $r_{\text{out}}$. As the specific energy of the flow remains constant, the thermal energy increases due to a drop in radial velocity at a given radius. Consequently, the electron temperature increases at that radius. Panels (e) and (f) show the respective mass density ($\rho$) profiles corresponding to Figs. \ref{fig:solutions}a-b. The flow density increases as we move toward the horizon. Moreover, as the critical points shift to smaller radii, $\rho$ increases. This behavior follows from the conservation of mass flux: as the velocity profile decreases, the density profile is expected to increase with the inward shift of the critical points toward the horizon. The profiles of the disc's aspect ratio ($H/r$), corresponding to the accretion solutions in Figs. \ref{fig:solutions}a–b, are shown in panels (g)–(h), respectively. It is observed that $H/r < 1$ throughout the disc. The relative thickness of the disc is small near the inner edge; however, it increases as we move away from the horizon. Additionally, $H/r$ increases with increasing $M_{\rm H}$ at a given $a_0$, or with decreasing $a_0$ at a given $M_{\rm H}$. This is naturally expected, as an increase in $M_{\rm H}$ or a decrease in $a_0$ raises the flow temperature, thereby increasing the thermal pressure, which ultimately leads to an enhancement of $H/r$. In Table~\ref{tab:table-1}, we summarize the properties of the critical points related to the accretion solutions presented in Figs.~\ref{fig:solutions}a-b. This table highlights the changing behaviors of the accretion solutions and illustrates the potential shifting of the critical points as the halo compactness increases, specifically in terms of increasing $M_{\rm H}$ and decreasing $a_0$. It is important to note that in \cite{Stuchlik-2021-059}, based on observed QPO frequencies around certain supermassive black holes, such as 1H0707-495, RE J1034+396, Mrk 766, and ESO 113-G010a, it is reported that $M_{\rm H}$ can have values from a few tenths to several hundred times $M_{\rm BH}$. Motivated by this phenomenological study, in our analysis, we consider $M_{\rm H}$ to be a few tenths of $M_{\rm BH}$, in order to ensure that $H/r < 1$ throughout the disc. A larger value of $M_{\rm H}$ leads to a higher flow temperature, and the corresponding $H/r$ can exceed unity, which violates the model assumptions. Nevertheless, there are other accretion models in the literature, such as the Novikov-Thorne model, where authors have considered $M_{\rm H}$ values up to a few hundred times $M_{\rm BH}$ \cite{Heydari-Fard-2025-49}. This is because, in the aforementioned thin disc model, the flow temperature is much lower than that in the hot accretion flow model.

\begin{figure*}
	\centering
	\includegraphics[width=0.975\linewidth]{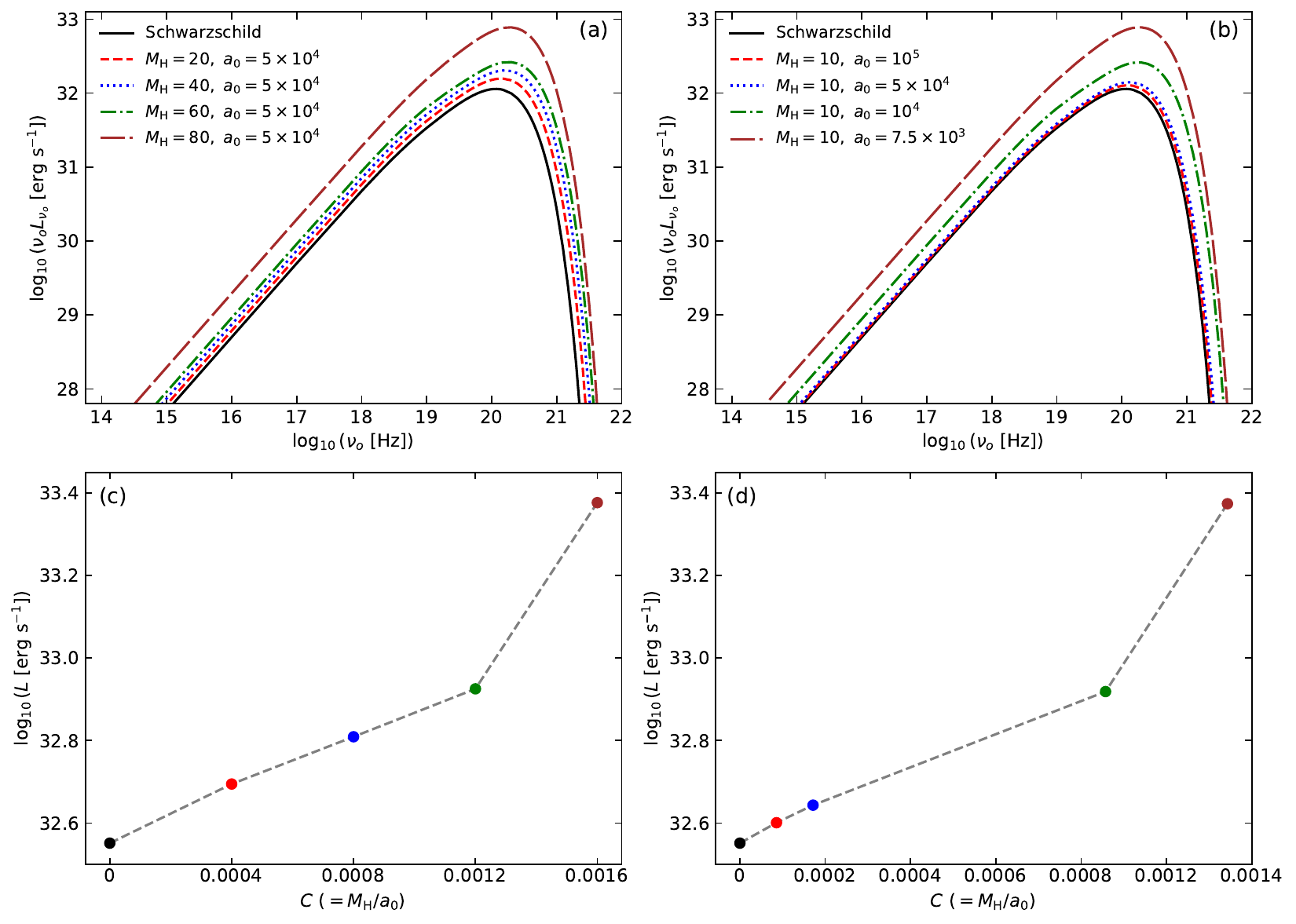}
	\caption{Panels (a) and (b) represent the spectral energy distribution (i.e., $\nu_o L_{\nu_o}$ versus $\nu_o$ curves) of the emitted radiation from the accretion disc for the global accretion solutions shown in Figs.~\ref{fig:solutions}a-b, respectively. The effect of halo compactness ($C$) on the bolometric disc luminosity ($L$) is depicted in panels (c) and (d). The filled circles represent the $L$ values corresponding to the global accretion solutions of Figs. \ref{fig:solutions}a-b. Here, the input parameters are chosen as $r_{\text{edge}} = 1000$, $\lambda = 3.1$, $E = 1.00025$, $M_{\text{BH}} = 10^6M_{\odot}$ and $\dot{M} = 10^{-5}\dot{M}_{\text{Edd}}$. See the text for details.}
	\label{fig:SED}
\end{figure*}

Next, we investigate the spectral properties of the accretion disc and examine how they are affected by the compactness of the dark matter halo. In this work, we consider a supermassive black hole with $M_{\text{BH}} = 10^6M_{\odot}$, where $M_{\odot}$ is the Solar mass. The mass accretion rate is taken to be very small as $\dot{M} = 10^{-5}\dot{M}_{\text{Edd}}$, where $\dot{M}_{\text{Edd}} = 1.39\times 10^{18} M_{\text{BH}}/M_{\odot}~\text{gm}~\text{s}^{-1}$ is the Eddington mass accretion rate. We calculate the spectral energy distribution (SED) associated with the global accretion solutions of Figs.~\ref{fig:solutions}a-b using Eq.~(\ref{eq:monochromatic-luminosity}). The obtained results are shown in the respective panels (a) and (b) of Fig.~\ref{fig:SED}, where the variation of the quantity $\nu_oL_{\nu_o}$ as a function of the observed frequency $\nu_o$ is depicted. In all cases, the emitted radiation maximizes power at $\nu_o \approx 10^{20}~\text{Hz}$. Also, the spectra exhibit a sharp cut-off around $\nu_o \approx 10^{22}~\text{Hz}$ ($= k_{\rm B}T_{e0}/h$), which corresponds to the disc inner edge electron temperature $T_{e0} \approx 10^{11}\text{K}$. We find that the SEDs for the Schwarzschild BH are lower than those for the Cardoso BH. This is because electron temperature across the entire disc in the Schwarzschild model is lower compared to the Cardoso model (see Figs.~\ref{fig:fp}c-d). Moreover, we observe that the SED increases with the rise in $M_{\rm H}$ and decrease in $a_0$, which is due to the corresponding increase in $T_e$, as shown in Figs.~\ref{fig:fp}c-d. Note that the above findings are consistent with the work in \cite{Heydari-Fard-2025-49}, where the authors studied the spectral properties of accretion flows by treating the disc as a perfect black body emitter. In that study, the hydrodynamics of the flow were governed by the geodesic equation of the particles, with the flow reaching up to the innermost stable circular orbit. Furthermore, the flow velocity never surpasses the local sound speed, implying that the transonic accretion model was not considered. Thereafter, we calculate the bolometric luminosity ($L$) of the accretion disc using Eq. (\ref{eq:bolometric-luminosity}) for the global accretion solutions of Figs. \ref{fig:solutions}a-b. The obtained results are depicted in panels (c) and (d) of Fig. \ref{fig:SED}, where the variation of $L$ as a function of compactness parameter ($C$) is shown. Here, the filled circles, using the same color codes as in Figs.~\ref{fig:solutions}a-b, joined by the dashed (gray) lines, denote the results for the respective accretion solutions. In both panels, we notice that $L$ increases with $C$. As the increase in $M_{\rm H}$ or decrease in $a_0$ enhances the SED (see Figs. \ref{fig:SED}a-b), it is therefore expected that $L$ (the area under the SED curve) will also increases with halo compactness. Such an increase in $L$ can also be directly followed from Eq. (\ref{eq:bolometric-luminosity}). As $C$ increases, $T_e$ rises (see Fig. \ref{fig:fp}c-d), which in turn increases $ L$, since it varies as $T_e^{1/2}$. A possible physical explanation for the increase of $L$ with $C$ is provided here. During accretion, as material moves toward smaller radii, it loses gravitational potential energy. A fraction of this lost potential energy can be emitted as electromagnetic radiation, making the disc luminous. When $M_{\rm H}$ increases or $a_0$ decreases, the density of the dark matter halo increases (see Eq. (\ref{eq:density-profile})). This leads to an increase in gravitational potential with the increase in halo compactness. Thus, a mass element of the fluid experiences larger gravitational potential energy. As a result, more potential energy is lost during accretion, which consequently enhances the disc luminosity.

\subsection{Accretion with shocks}
\label{sec:shock}

\begin{figure}
	\centering
	\includegraphics[width=\columnwidth]{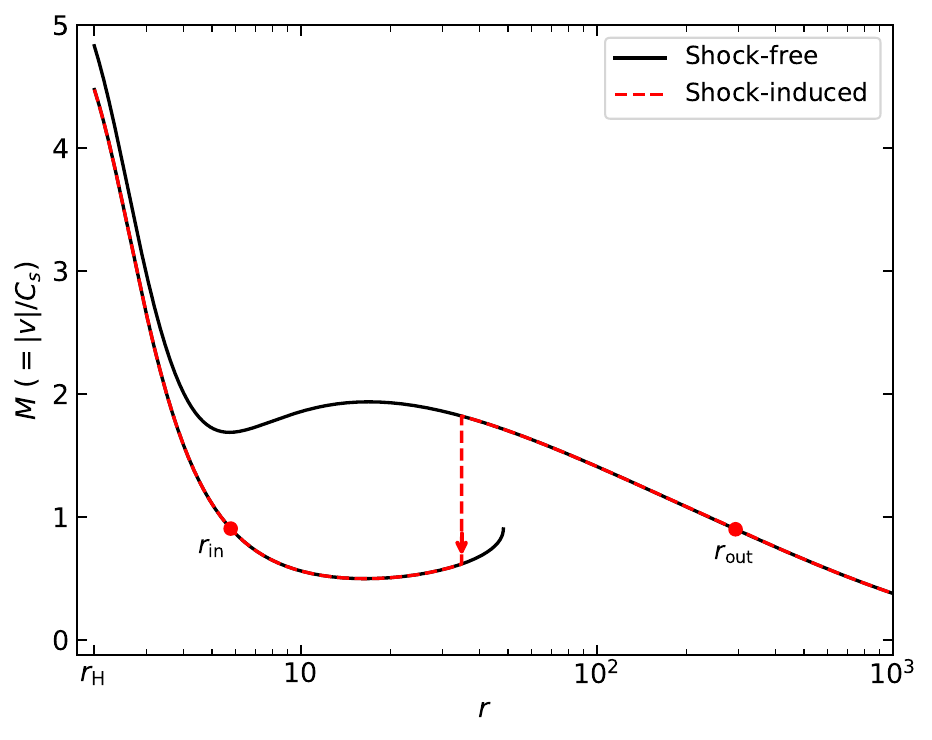}
	\caption{Accretion solutions for the shock-free and shock-induced scenarios. The vertical line indicates the shock location ($r_{\text{sh}}$) and the arrow represents the overall direction of the flow. In this figure, we choose $\lambda = 3.1$, $E = 1.0005$, $M_{\rm H} = 10$, and $a_0 = 10^5$. See the text for details.}
	\label{fig:shock-solution}
\end{figure}

We previously mentioned that A-type solution topology can accommodate shock transitions, depending on several physical conditions. In this section, we illustrate such shock scenarios and analyze their characteristics in terms of flow parameters. Fig.~\ref{fig:shock-solution} shows a typical A-type solution topology (solid black lines) for the set of global constants $(\lambda, E, M_{\rm H}, a_0) = (3.1, 1.0005, 10, 10^5)$. Here, the global solution passes through $r_{\text{out}} = 293.7518$, and, the closed solution, passing through $r_{\text{in}} = 5.7833$, is truncated at a radius $r_{t} = 48.1933$. The entropy accretion rates ($\mathcal{\dot{M}}$) for the inner and outer branches are calculated to be $3.3296 \times 10^7$ and $2.1058 \times 10^7$, respectively. Since the inner solution has a higher entropy content than the outer solution, the flow prefers to jump into the inner closed branch in the form of standing shocks, provided the relativistic shock conditions are satisfied. We calculate the shock location ($r_{\text{sh}}$) using the Rankine-Hugoniot standing shock conditions as \cite{Taub-1948-328},
\begin{align}
\label{eq:shock-condition-1}
& \left[\rho u^r\right] = 0,\\
\label{eq:shock-condition-2}
& \left[(e + p)u^ru^t\right] = 0,\\
\label{eq:shock-condition-3}
& \left[(e + p)u^ru^r + pg^{rr}\right] = 0,
\end{align}
where the square brackets denote the difference of the quantities across $r_{\text{sh}}$. Eqs.~(\ref{eq:shock-condition-1}), (\ref{eq:shock-condition-2}), and (\ref{eq:shock-condition-3}) correspond to the conservation of mass flux, energy flux, and radial-momentum flux across $r_{\text{sh}}$, respectively. The dashed (red) curve in Fig.~\ref{fig:shock-solution} illustrates a shock-induced accretion solution, with a noticeable sharp jump at $r_{\text{sh}} = 34.8633$. Note that the shock solutions can pass through both $r_{\text{in}}$ and $r_{\text{out}}$ simultaneously.

\begin{figure*}
	\centering
	\includegraphics[width=0.75\linewidth]{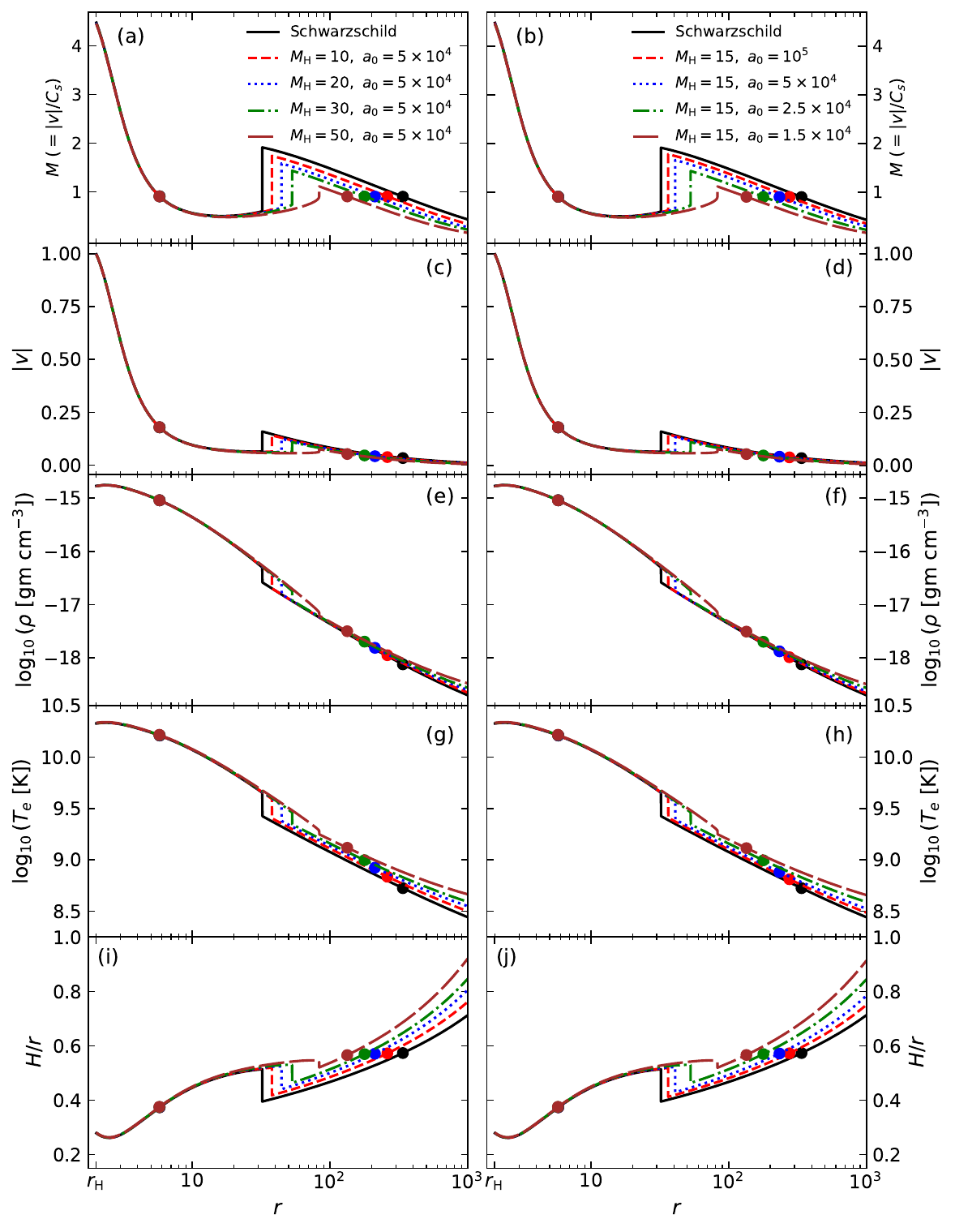}
	\caption{Shock solutions for halo masses $M_{\rm H} = 10$, $20$, $30$, and $50$ with length scale $a_0 = 5 \times 10^4$ (panel (a)), and for $a_0 = 10^5$, $5 \times 10^4$, $2.5 \times 10^4$, and $1.5 \times 10^4$ with $M_{\rm H} = 15$ (panel (b)). The corresponding radial velocity ($v$), mass density ($\rho$), electron temperature ($T_e$), and aspect ratio ($H/r$) profiles associated with these shock solutions are presented in panels (c)-(d), (e)-(f), (g)-(h), and (i)-(j), respectively. In each panel, the shock locations are marked by vertical lines. In this figure, we set $\lambda = 3.1$, $E = 1.0005$, $M_{\text{BH}} = 10^{6}M_{\odot}$, and $\dot{M} = 10^{-5}\dot{M}_{\text{Edd}}$. See the text for details.}
	\label{fig:shock-solutions-diff=C}
\end{figure*}

We now explore various shock properties (e.g., shock radius, density compression, and temperature compression at shock location, etc.) in the presence of a dark matter halo and compare them with those for a Schwarzschild BH. In panels (a) and (b) of Fig.~\ref{fig:shock-solutions-diff=C}, we present the shock solutions for different values of halo mass $M_{\rm H}$ with a fixed length scale $a_0 = 5 \times 10^4$, and for various values of $a_0$ with $M_{\rm H} = 15$. In this case, a given set of flow parameters is chosen as $(\lambda, E) = (3.1, 1.0005)$. Here, the solid (black) curves represent the result for the Scwarzschild BH. The dashed (red), dotted (blue), dash-dotted (green), and long-dashed (brown) lines correspond to the Cardoso BH with $M_{\rm H} = 10$, $20$, $30$, and $50$, respectively. Similar color codes are used for $a_0 = 10^5$, $5 \times 10^4$, $2.5 \times 10^4$, and $1.5 \times 10^4$, respectively. Using the shock conditions (\ref{eq:shock-condition-1}), (\ref{eq:shock-condition-2}), and (\ref{eq:shock-condition-3}), we obtain the shock locations at $r_{\text{sh}} = 32.1569$, $37.7497$, $44.3725$, $52.9654$, and $83.3215$ for $M_{\rm H} = 0$, $10$, $20$, $30$, and $50$, respectively. On the other hand, for $a_0 = 10^5$, $5 \times 10^4$, $2.5 \times 10^4$, and $1.5 \times 10^4$, the calculated shock locations are $r_{\text{sh}} = 36.2815$, $40.8861$, $52.8754$, and $82.8315$, respectively. The critical points and shock locations associated with these solutions are summarized in Table \ref{tab:table-2}. It is observed that the shock locations for the Schwarzschild BH are located closer to the horizon compared to those for the Cardoso BH. Furthermore, as the compactness of the halo increases, the shock fronts move away from the central object. For these shock solutions, the profiles of radial velocity ($v$), mass density ($\rho$), electron temperature ($T_e$), and aspect ratio ($H/r$) are presented in Figs.~\ref{fig:shock-solutions-diff=C}c-d, \ref{fig:shock-solutions-diff=C}e-f, \ref{fig:shock-solutions-diff=C}g-h, and \ref{fig:shock-solutions-diff=C}i-j, respectively. It is observed that the analyzed flow variables undergo significant changes across the shock fronts. This occurs because, according to the shock condition (\ref{eq:shock-condition-1}), as $v$ decreases at $r_{\text{sh}}$, $\rho$ increases. Also, due to the drop in $v$, kinetic energy of the flow is converted into thermal energy, resulting in an increase in $T_e$ at $r_{\text{sh}}$. Consequently, $H/r$ increases at $r_{\rm sh}$, but it continues to remain below unity. Moreover, we observe that the change in $v$ at $r_{\text{sh}}$ diminishes as the shock originates at larger radii, decreasing the difference of $\rho$, $T_e$, and $H/r$ across the shock fronts. We wish to mention that $r_{\text{sh}}$ determines the size of the PSC, where a swarm of hot electrons can produce high-energy radiation through inverse Compton scattering. Such emissions are commonly observed in AGNs \cite{Majumder-2022-2508, Nandi-2024-1149, Chatterjee-2024-148}. Furthermore, the oscillation of PSC can lead to QPOs in their power density spectra \cite{molteni-1995, Dihingia-2019-2412, Patra-2024-371}. When $r_{\rm sh}$ shifts, it directly impacts the QPO frequencies. Therefore, from the observed QPO frequency in the AGN spectrum, it may be possible to determine the presence of a dark matter halo. Even though the study of QPOs is beyond the scope of this paper, we provide a brief discussion on this topic. Numerical simulations have shown that when the infall timescale from the shock location is comparable to the post-shock cooling timescale, the shock can oscillate in a quasi-periodic manner \cite{molteni-1995}, where the QPO frequency is given by $\nu_{\rm QPO} = 1/t_{\rm infall}$. Here, $t_{\rm infall}$ is the infall time of the flow from $r_{\rm sh}$ to $r_{\rm H}$, i.e., $t_{\rm infall} = \int_{r_{\rm sh}}^{r_{\rm H}}dt = \int_{r_{\rm sh}}^{r_{\rm H}}\frac{1}{v(r)}dr$, where $v(r)$ is the post-shock velocity of the flow. This equation clearly indicates that when $r_{\rm sh}$ shifts to a larger radius due to the presence of dark matter halo, the post-shock flow takes a longer time to reach the horizon, i.e., $t_{\rm infall}$ increases. As a result, $\nu_{\rm QPO}$ decreases compared to the usual Schwarzschild BH, where $r_{\rm sh}$ forms at smaller radii than in the Cardoso BH case.

\begin{table*}
	\centering
	\caption{Dark matter halo mass ($M_{\rm H}$), halo length scale ($a_0$), critical point locations ($r_{\text{in}}, r_{\text{out}}$), and shock location ($r_{\text{sh}}$) for the shock solutions presented in Fig.~\ref{fig:shock-solutions-diff=C}.}
	\label{tab:table-2}
	\begin{ruledtabular}
		\begin{tabular}{lcccc}
			$M_{\rm H}$ & $a_0$ & $r_{\text{in}}$ & $r_{\text{out}}$ & $r_{\text{sh}}$\\
			\hline
			$0$ (Schwarzschild) & --- & $5.7869$ & $337.7180$ & $32.1569$\\
			$10$ & $5 \times 10^4$ & $5.7797$ & $260.3761$ & $37.7497$\\
			$20$ & $5 \times 10^4$ & $5.7725$ & $211.6646$ & $44.3725$\\
			$30$ & $5 \times 10^4$ & $5.7654$ & $177.7814$ & $52.9654$\\
			$50$ & $5 \times 10^4$ & $5.7515$ & $133.1758$ & $83.3215$\\
			$15$ & $10^5$ & $5.7815$ & $275.8587$ & $36.2815$\\
			$15$ & $5\times 10^4$ & $5.7761$ & $233.5852$ & $40.8861$\\
			$15$ & $2.5 \times 10^4$ & $5.7654$ & $178.4825$ & $52.8754$\\
			$15$ & $1.5 \times 10^4$ & $5.7515$ & $134.4154$ & $82.8315$\\
		\end{tabular}
	\end{ruledtabular}
\end{table*}

\subsection{Shock parameter space}
\label{sec:shock-parameter-space}

\begin{figure*}
	\centering
	\includegraphics[width=\linewidth]{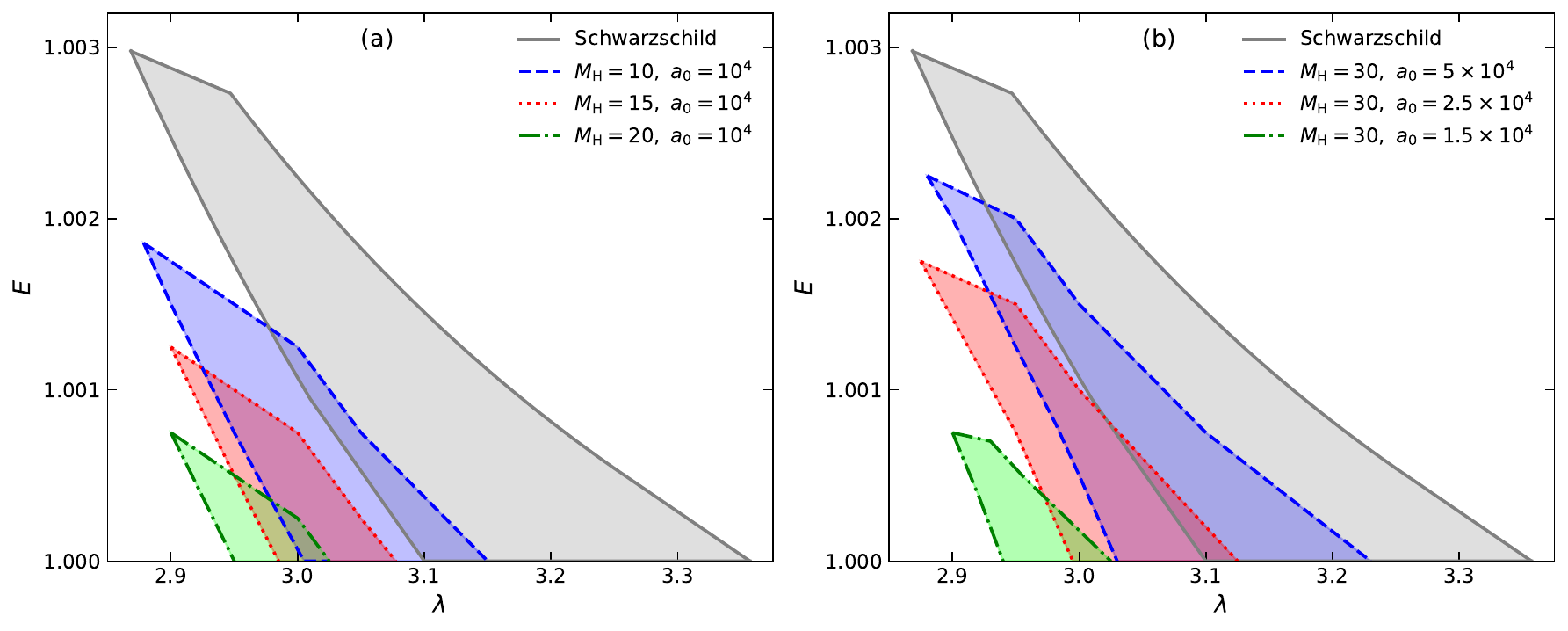}
	\caption{Modification of the shock parameter space in specific angular momentum ($\lambda$) and energy ($E$) plane for the halo masses $M_{\text{H}} = 10$, $15$, and $20$ with $a_0 = 10^4$ (panel (a)), and for $a_0 = 5 \times 10^4$, $2.5 \times 10^4$, and $1.5 \times 10^4$ with $M_{\rm H} = 30$ (panel (b)). In each panel, the effective region within the gray (solid) line corresponds to the Schwarzschild BH without a dark matter halo (i.e., $M_{\rm H} = 0$). See the text for details.}
	\label{fig:shock-parameter-space}
\end{figure*}

Here, we identify the effective region of the specific angular momentum ($\lambda$) and energy ($E$) that admits shock solutions for the Cardoso BH and compare it with that of the Schwarzschild BH. In Fig.~\ref{fig:shock-parameter-space}a, we present the shock parameter space in the $\lambda - E$ plane for different values of halo mass $M_{\text{H}} = 10$, $15$, and $20$ with a fixed halo length scale of $a_0 = 10^4$. The regions bounded by the dashed (blue), dotted (red), and dash-dotted (green) curves represent the results for $M_{\rm H} = 10$, $15$, and $20$, respectively. The parameter space enclosed by the solid (gray) line corresponds to the Schwarzschild BH model. We observe that for the Schwarzschild BH, flow exhibits shocks at relatively higher $\lambda$ and $E$ values compared to the Cardoso BH. Also, the area under the shock parameter space is larger for the Schwarzschild BH than for the Cardoso BH. As $M_{\rm H}$ increases, the parameter space shifts toward lower $\lambda$ and $E$ domains, and the parameter space gradually shrinks as well. Similarly, in Fig.~\ref{fig:shock-parameter-space}b, we present the modification of the shock parameter space in the $\lambda-E$ plane for varying $a_0$ with a fixed $M_{\rm H} = 30$. The dashed (blue), dotted (red), and dash-dotted (green) curves correspond to $a_0 = 5 \times 10^4$, $2.5 \times 10^4$, and $1.5 \times 10^4$, respectively. The shock parameter space for the Schwarzschild BH is bounded by the solid (gray) line. As in Fig.~\ref{fig:shock-parameter-space}a, we find that the shock parameter space for the Schwarzschild BH without dark matter halo can accommodate higher $\lambda$ and $E$ values than in the presence of a halo. Moreover, the area under the parameter space decreases gradually when $a_0$ decreases. 

\section{Conclusion and discussion}
\label{sec:conclusion}

In this work, we explore the transonic accretion flow around a galactic black hole with a dark matter halo, as proposed in \cite{Cardoso-2022-L061501}. The flow hydrodynamics in the accretion disc are modeled within fully general relativistic framework. Using the relativistic equation of state, we numerically solve the radial momentum and energy equations. Consequently, we obtain the global accretion solutions in both the presence and absence of shocks. The main objective of this work is to explore the effect of halo mass ($M_{\rm H}$) and length scale ($a_0$) on the physical properties of the accretion disc. We make an effort to compare these results with those for the usual Schwarzschild BH without a dark matter halo. We summarize our findings point-wise below. 

\begin{itemize}	
\item{We find A and W-type accretion solution topologies, where the flow possesses multiple critical points. We observe that for the Schwarzschild BH and the Cardoso BH with small halo compactness, the solution topology remains A-type. However, at higher $M_{\rm H}$ or lower $a_0$ values, the solution topology changes to W-type.}

\item{We also observe that A-type solution topologies can admit standing shock transitions when the flow satisfies the relativistic shock conditions. It is noticed that the shock solutions are not unique but rather exist within a broad range of the parameter space spanned by the flow specific angular momentum ($\lambda$) and energy ($E$). Accordingly, we examine the modification of the shock parameter space as a function of $M_{\rm H}$ and $a_0$.  We observe that the shock parameter space is larger for the Schwarzschild BH compared to the Cardoso BH. Also, an increase in the halo compactness shrinks the shock parameter space towards lower ($\lambda, E$) domain. Therefore, the shock solutions exist in an extremely small $\lambda - E$ parameter spacetime for high halo compactness.}

\item{Furthermore, we examine the effect of $M_{\rm H}$ and $a_0$ on various shock properties, particularly the shock location ($r_{\text{sh}}$), as well as the changes in mass density ($\rho$) and electron temperature ($T_e$) across the shock fronts. We find that the shock fronts settle down at larger radii for the Cardoso BH compared to the Schwarzschild BH. Moreover, as $M_{\rm H}$ increases or $a_0$ decreases, $r_{\text{sh}}$ moves away from the horizon, leading to a decrease in the changes of $\rho $ and $T_e$ across $r_{\text{sh}}$. When the compactness of the dark matter halo is high, significant deviations in the shock properties are observed compared to the Schwarzschild BH model.}

\item{In addition, we calculate the spectral energy distribution (SED) for the global accretion solutions in the A and W-type solution topologies using the relativistic thermal bremsstrahlung emission coefficient. We find that the SED increases with increasing $M_{\rm_H}$ or decreasing $a_0$. When the halo compactness is low, the SEDs for the Cardoso BH differ barely from those of the Schwarzschild BH. However, a noticeable difference between the SEDs for these two black hole models emerges when the halo compactness is high. These results are consistent with existing work in the literature based on a different accretion model \cite{Heydari-Fard-2025-49}. While investigating the bolometric disc luminosity ($L$), we observe that $L$ increases with halo compactness. A considerable change in $L$ is evident for high halo compactness compared to that of the Schwarzschild BH model. Therefore, such quantitative variations in the SED and $L$ provide a clear distinction between the Cardoso BH and Schwarzschild BH models.} 	
\end{itemize}

Here, we would like to mention that in our analysis, we have chosen a very small mass accretion rate, $\dot{M} = 10^{-5}\dot{M}_{\rm Edd}$. But what happens when a higher value of $\dot{M}$ is considered? Would it wash out the effect of the dark matter halo? We now shed light on this question. The two fundamental parameters, $T_e$ and $v$, are calculated from Eqs. (\ref{eq:temperature-gradient}) and (\ref{eq:velocity-gradient}), respectively, which are derived by imposing $d\dot{M}/dr = 0$ (i.e., $\dot{M} = \text{constant}$). Therefore, for a given accretion solution, the profiles of $T_e$ and $v$ do not change with an increase in $\dot{M}$. However, since $\rho$ is computed using the expression for $\dot{M}$ (Eq. (\ref{eq:mass-accretion-rate})), its profile scales to higher values accordingly. As a result, the SED and $L$, which depend on $T_e$, v, and $\rho$, also scale to higher values with increasing $\dot{M}$. This clearly indicates that increasing $\dot{M}$, simply shifts all the results in Fig. \ref{fig:SED} to higher values by the equal factor. Also, the scaling of $\rho$ does not affect $r_{\rm sh}$, because it is determined using the shock conditions (Eqs. (\ref{eq:shock-condition-1}), (\ref{eq:shock-condition-2}), and (\ref{eq:shock-condition-3})), which involve only the differences of the respective quantities at a given radius. Thus, the QPO frequency, which depends on $r_{\rm sh}$ and $v$, is also not affected. Hence, in our model, the effect of dark matter is not washed out when a larger $\dot{M}$ is considered. In that sense, the chosen value of $\dot{M}$ has no such direct connection to the enhancement of the dark matter effect in the identified results.

In conclusion, this study suggests that in the presence of a dark matter halo with compactness typically $C < 5 \times 10^{-4}$, the results are nearly indistinguishable from those of an isolated Schwarzschild black hole. As $C$ increases beyond this limit, we observe a noticeable deviation from the Schwarzschild black hole model. For higher halo compactness, roughly $C \gtrsim 10^{-3}$, the profiles of various thermodynamic variables (e.g., electron temperature, mass density, etc.) associated with the accretion solutions become more distinguishable compared to the Schwarzschild case. As a result, spectral properties such as the disc luminosity distribution and bolometric luminosity are modified by the presence of the dark matter halo, compared to the vacuum Schwarzschild black hole model. So, as $C$ exceeds approximately $5 \times 10^{-4}$, deviations from the Schwarzschild model begin to occur and become more prominent with increasing $C$. However, such an increase in $C$ is constrained by our model assumptions, specifically the validity of the thin disc model, which requires $H/r < 1$. For shock solutions, various shock properties (e.g., shock radius, density compression, temperature compression across the shock radius, etc.) exhibit potential imprints of the presence of a dark matter halo with high compactness. However, even in this highly compact regime, the electron temperature and mass density profiles do not vary significantly with compactness compared to the Schwarzschild black hole case; therefore, we expect negligible changes in the corresponding SEDs. On the other hand, we expect the QPO frequency to undergo a significant change due to the substantial shift in the shock location when a high compactness halo is considered.

As discussed in the previous paragraph, the accretion features of Cardoso BHs differ from those of Schwarzschild BHs at high halo compactness. However, an obvious question arises regarding the plausibility of achieving a high-compactness halos in real astrophysical settings. In the Cardoso model, the density of the dark matter halo gradually increases toward the galaxy's core and vanishes at the event horizon of the central black hole \cite{Cardoso-2022-L061501, Chowdhury-2025-08528, Heydari-Fard-2025-49}. Essentially, a cuspy density profile is found at small scales near the galactic center. While such cuspy profiles may arise in massive galaxies, they contradict the observed cored (or flat) density profiles commonly found in dwarf galaxies \cite[and referrences therein]{Eymeren-2009-1, Dekel-2017-1005, Relatores-2019-94, Almeida-2020-L14, Lazar-2020-2393}. This cusp-core discrepancy remains an ongoing topic of debate within the astrophysics community. Since high compactness parameters may lead to steeper cuspy profiles in the Cardoso model, the corresponding accretion properties require further verification using a more realistic black hole model that aligns with various observational phenomena. Thereafter, we can better assess whether highly compact halos can truly exist. At this stage, our predictions related to high halo compactness should be taken as suggestive. To the best of our knowledge, all existing black hole metrics in the literature suffer from this cusp problem. Nevertheless, we anticipate that a more realistic metric may be developed in the future, which could enable more definitive answer.

Finally, we highlight the limitations of our work. In our model, the angular momentum is a conserved quantity due to the assumption of ideal fluid dynamics. However, the viscous stress can transport angular momentum towards the outer edge of the disc \cite{Dihingia-2019-2412, Singh-2023-1, Patra-2024-371}. Additionally, our accretion model does not account for magnetic fields. The presence of large-scale magnetic fields can alter the dynamics of the accreting material through the magnetic pressure \cite{Mitra-2022-5092, Mitra-2024-28}. Also, due to the presence of magnetic fields, electrons in the hot plasma can emit synchrotron radiation in addition to thermal bremsstrahlung radiation. However, in this study, we focus solely on the bremsstrahlung component when calculating the SED, as the accretion model does not include magnetic fields. Furthermore, we do not account for other emission processes, such as Comptonization \cite{Mahadevan-1997-585, Witzel-2018-15, Dihingia-2019-3043, Sarkar-2022-34}. If all these processes were incorporated into the total emission coefficient, the shape of the resulting SED would change. Typically, it would exhibit two peaks: one in the radio or near-infrared range (due to the synchrotron component) and another in the hard X-ray range (due to the bremsstrahlung component, similar to our case). A model spectrum that includes all of these components is essential for analyzing the observed AGNs and BH-XRBs spectra. Moreover, due to the high temperature gradient near the inner regions of the disc, thermal conduction plays an important role in influencing the behavior of the accretion disc \cite{Mitra-2023-4431, Singh-2024-068}. However, the present study has neglected the thermal conduction. In addition, we adopt a simple scaling relation between the electron and ion temperatures, although several studies in the literature have explored two-temperature accretion flows \cite{Sarkar-2018-1950037, Dihingia-2018-1, Dihingia-2020-3043, Sarkar-2020-A209, Sarkar-2022-34}. It is important to note that these physical processes are highly relevant in the context of black hole accretion flows. We plan to address these aspects in future work and report the outcomes elsewhere.  

\section*{Data availability statement}
The data underlying this article will be available with reasonable request.

\section*{Acknowledgments}
The authors thank the anonymous reviewer for constructive comments and useful suggestions that helped to improve the quality of the paper significantly. The authors would also like to thanks Chiranjeeb Singha, Soumya Bhattacharya and Samik Mitra for useful discussions. SP acknowledges the University Grants Commission (UGC), India, for the financial support through the Senior Research Fellowship (SRF) scheme. The work of BRM is supported by a START-UP RESEARCH GRANT from the Indian Institute of Technology Guwahati (IIT Guwahati), India, under the grant SG/PHY/P/BRM/01.
 
\bibliographystyle{apsrev}
\bibliography{references-CBH.bib}

\end{document}